\documentclass[prc,twocolumn,superscriptaddress,nofootinbib]{revtex4-1}

\usepackage{newtxtext}
\usepackage[varvw,bigdelims]{newtxmath}

\usepackage[colorlinks=true,allcolors=blue]{hyperref}
\usepackage{graphicx}
\usepackage{bm}
\usepackage{amsmath}
\begin{document}
\title{Model independent analysis of femtoscopic correlation functions: \\An application to the $D_{s0}^*(2317)$}

\author{Natsumi Ikeno}
\email{ikeno@tottori-u.ac.jp}
\affiliation{Department of Agricultural, Life and Environmental Sciences, Tottori University, Tottori 680-8551, Japan}
\affiliation{Cyclotron Institute, Texas A\&M University, College Station, Texas 77843, USA}

\author{Genaro Toledo}
\email{toledo@fisica.unam.mx}
\affiliation{Instituto de F\'{i}sica, Universidad Nacional Autonoma de Mexico, AP20-364, Ciudad de Mexico 01000, Mexico}

\author{Eulogio Oset}
\email{oset@ific.uv.es}
\affiliation{Departamento de F\'{i}sica Te\'{o}rica and IFIC, Centro Mixto Universidad de Valencia - CSIC,
Institutos de Investigaci\'{o}n de Paterna, Aptdo. 22085, 46071 Valencia, Spain}

\preprint{}

\date{\today}

\begin{abstract}
We face the inverse problem of obtaining the interaction between coupled channels from the correlation functions of these channels. We apply the method to the interaction of the $D^0 K^+$, $D^+ K^0$, and $D^+_s \eta$ channels, from where the $D^*_{s0}(2317)$ state emerges. We use synthetic data extracted from an interaction model based on the local hidden gauge approach and find that the inverse problem can determine the existence of a bound state of the system with a precision of about 20 MeV. At the same time, we can determine the isospin nature of the bound state and its compositeness in terms of the channels. Furthermore, we evaluate the scattering length and effective range of all three channels, as well as the couplings of the bound state found to all the components. Lastly, the size parameter of the source function, $R$, which in principle should be a magnitude provided by the experimental teams, can be obtained from a fit to the data with relatively high accuracy. 
These findings show the value of the correlation function to learn about the meson–meson interaction for systems which are difficult to access in other present facilities.
\end{abstract}

%\pacs{14.40.Rt,12.40.Yx, 13.75.Lb}

\maketitle

%%%%%%%%%%%%%%%%%%%%%%%%%%%%%%%%%%%%%%%%%%%%%%%%%%%%%%%%%%%%%%%%%%%%%%%%%%%%%%%%%% 
\section{Introduction}
The femtoscopic correlations function measurements \cite{STAR:2014dcy,ALICE:2017jto,STAR:2018uho,ALICE:2018ysd,ALICE:2019hdt,ALICE:2019eol,ALICE:2019buq,ALICE:2019gcn,ALICE:2020mfd,ALICE:2021szj,ALICE:2021cpv,Fabbietti:2020bfg,ALICE:2022enj} have opened a new window to learn about hadron interactions and much theoretical work is following that path~\cite{Morita:2014kza,Ohnishi:2016elb,Morita:2016auo,Hatsuda:2017uxk,Mihaylov:2018rva,Haidenbauer:2018jvl,Morita:2019rph,Kamiya:2019uiw,Kamiya:2021hdb,Kamiya:2022thy,Liu:2023uly,Vidana:2023olz,Albaladejo:2023pzq}. The correlation functions provide information on the wave function of the pairs of particles observed in the experiment, from where one can obtain information on the interaction of these particles. The method will prove complementary to the studies done by other facilities such as LHCb, Belle, and BESIII. On the first hand, there is in principle no limitation to the pairs that one selects, and hence one can have access to the interaction of many pairs which would not be accessible with other facilities. On the second hand, the experiments select pairs of particles and the correlation functions are sensitive to the interaction of the pairs. Conversely, in the most popular and successful methods, one studies three body decays of some hadrons, let's say $B$ decays, then one can play around the invariant mass of different pairs, but the analysis of the data are complicated, and sometimes ambiguous, because one has to deal simultaneously with the interaction of these pairs, and resonances appearing in some channel produce reflections in other channels that one must be careful not to identify with new resonances \cite{LHCb:2022bkt,BESIII:2021aza,Song:2022kac}.

So far, in order to learn about the interaction of particles from the correlation functions, one compares experimental correlation functions with models for the interaction, with the aim of discriminating among models, or allowing the determination of certain parameters from a given model~\cite{Kamiya:2022thy,Kamiya:2019uiw}. In the present work, we follow a different path and use a method that allows one to determine the physical magnitudes related to the studied channel, eventually determining if there are some bound states below the threshold, as well as to determine the nature of the states obtained. The method starts considering the most general interaction potential in coupled channels, which involves a few parameters. Then the correlation functions of coupled channels are evaluated and a simultaneous fit to these correlation functions is conducted to determine the parameters of the theory. We realize that there are large correlations between the parameters and we play with the freedom to select them in order to quantify uncertainties in the magnitudes evaluated. The method also allows for the contribution of missing channels or the existence of a state of nonmolecular nature that can couple to these channels. These possible contributions, with their uncertainty, also come from a fit to the correlation functions, such that in the case of the presence of bound states related to the studied channels, we can determine the probability that the state is of molecular nature formed by these channels and their relative contributions to the full wave function of the state.

In the present work, we apply the method to the $D^0 K^+$, $D^+ K^0$, and $D^+_s \eta$ channels as an example, and we show that by looking at the correlation function of these three channels, we can deduce that there is a state bound by about 42~MeV, that can be associated to be $D^*_{s0}(2317)$ state, and we can say something about the nature of this state. In the absence of real data, we produce synthetic data from a successful model. The inverse method used here allows us to obtain information from the original model, which is sufficient to determine physical quantities and tell us with which precision we can obtain them given the information provided by the correlation functions.

%%%%%%%%%%%%%%%%%%%%%%%%%%%%%%%%%%%%%%%%%%%%%%%%%%%%%%%%%%%%%%%%%%%%%%%%%%%%%%%%%%% 
\section{Formalism of the inverse problem for the determination of the $D^*_{s0}(2317)$}
The $D^*_{s0}(2317)$ is supposed to be a state of molecular nature from the channels $D^0 K^+$, $D^+ K^0$, and $D^+_s \eta$ \cite{vanBeveren:2003kd,Barnes:2003dj,Chen:2004dy,Kolomeitsev:2003ac,Gamermann:2006nm,Guo:2006rp,Yang:2021tvc,Liu:2022dmm}, mostly a  $D K$ state in isospin $I=0$ which is also supported by lattice QCD simulations~\cite{Mohler:2013rwa,Lang:2014yfa,Bali:2017pdv,Cheung:2020mql}. A detailed analysis of lattice QCD data done in Ref.~\cite{MartinezTorres:2014kpc} could even qualify the content of $DK$ in the wave function of the $D^*_{s0}(2317)$ around 72\% probability. The state is bound by about 42~MeV with respect to the $D^0 K^+$ threshold, a large binding compared to 360~keV of the $T_{cc}(3875)$ with respect to the $D^0 D^{*+}$ threshold, which is assumed to be a bound state of the $D^0 D^{*+}$ and $D^* D^{*0}$ channels~\cite{LHCb:2021auc} (see also this conclusion in the analysis of Ref.~\cite{Dai:2023cyo}). The state is observed in the $D^+_s \pi^0$ channel~\cite{BaBar:2003oey}, which violates isospin, and  this is the reason for its extremely small width ($< 3.8$~MeV).

In the present work, we shall assume that three correlation functions of the $D^0 K^+$, $D^+ K^0$ and $D^+_s \eta$ are given to us, and from there we shall try to induce the existence of the $D^*_{s0}(2317)$ state and its properties. These correlation functions are not yet available, but plans are already made to extend the study of the correlation functions to the $D$ sector in a near future~\cite{ALICE:2022wwr}. Then, to explain how the method works and what it accomplishes, we shall generate the correlation function using an updated version of the work of Ref.~\cite{Liu:2023uly}.

\subsection{Model for the $D^0 K^+$, $D^+ K^0$,  $D^+_s \eta$ interaction }\label{sec:model}
\begin{figure}[!t]
\begin{center}
\includegraphics[width=0.8\linewidth]{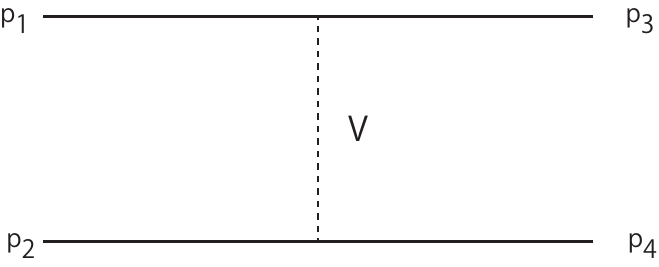}
\caption{Diagrammatic representation of the vector exchange that generates the interaction.}
\label{fig:interact}
\end{center}
\end{figure}

We use an extension of the local hidden gauge approach \cite{Bando:1987br,Harada:2003jx,Meissner:1987ge,Nagahiro:2008cv} in which the interaction is obtained from the exchange of vector mesons as depicted in Fig.~\ref{fig:interact}.

The interaction between the channels $D^0 K^+$(1), $D^+ K^0$(2),  $D^+_s \eta$(3) is given by
\begin{align}\label{eq:Vpot}
& V_{ij} = C_{ij} \ g^2 (p_1+p_3)\cdot (p_2+p_4) ;  \\ 
& g = \frac{M_V}{2f}, \   M_V =800~\text{MeV}, \  f=93~\text{MeV}, \nonumber
\end{align}
where
\begin{eqnarray}
C_{ij}= \left(
           \begin{array}{ccc}
            -\frac{1}{2}\left( \frac{1}{M_\rho^2} + \frac{1}{M_\omega^2} \right) & -\frac{1}{M_\rho^2} & \frac{2}{\sqrt{3}} \frac{1}{M^2_{K^*}}  \\[0.1cm]
           & -\frac{1}{2}\left( \frac{1}{M_\rho^2} + \frac{1}{M_\omega^2} \right)  & \frac{2}{\sqrt{3}} \frac{1}{M^2_{K^*}}  \\ 
            & & 0\
           \end{array}
         \right) \,,
\end{eqnarray}
and projected over $s$-wave
\begin{align}\label{eq:p1p3p2p4}
(p_1+p_3) \cdot (p_2+p_4)&\to \frac{1}{2}\big[3s-(M^2+m^2+{M'}^2+{m'}^2) \nonumber\\
&-\frac{1}{s}(M^2-m^2) ({M'}^2-{m'}^2)\big]\,,
\end{align}
where $M$, $m$ are the masses of the $D$, $K$ initial mesons in Fig.~\ref{fig:interact} and $M'$, $m'$ the corresponding ones for the final state. With the usual phase convention for the isospin doublet $(D^+, -D^0)$, $(K^+, K^0)$, the isospin states are given by 
\begin{equation}
\begin{aligned}
  |D K, I=0\rangle &= \frac{1}{\sqrt{2}} (D^{+} K^0 + D^{0} K^+)\,,  \\
  |D K, I=1, I_3 =0\rangle &= \frac{1}{\sqrt{2}} (D^{+} K^0 - D^{0} K^+)\,.
\label{eq:iso}
\end{aligned}
\end{equation}
We immediately observe that for the isospin $I=0$ the $C_{ij}$ coefficient becomes $C_{00} = C_{11} + C_{12} \simeq \frac{-2}{M^2_\rho}$ while for $I=1$ we have $C_{11} \simeq 0$, taking the same mass for the $\rho$ and $\omega$. Thus, we have attraction for  $I=0$, while the interaction is null for $I=1$ and we get a bound state for $I=0$ corresponding to the  $D^*_{s0}(2317)$. The bound state is found from the pole in the scattering matrix
\begin{equation}
T = [1-VG]^{-1} V\, ,
\label{eq:TBS}
\end{equation}
where $G$ is the loop function of the $DK$ or $D^+_s \eta$ states, $G= \text{diag}(G_i)$, which we regularize with a sharp cut off as
\begin{eqnarray}\label{eq:Gcut}
G_i (s) = \int_{|{\bm q}| <q_{\rm max}} \frac{d^3q}{(2\pi)^3} \, \frac{\omega_1 + \omega_2}{2 \,\omega_1  \omega_2} \,\frac{1}{s-(\omega_1 + \omega_2)^2+i\epsilon} 
\end{eqnarray}
with $\omega_1 = \sqrt{M^2 + {\bm{q}}^2 }$, $\omega_2 = \sqrt{m^2 + {\bm{q}}^2 }$ for each of the $i$ channels $D^0 K^+$, $D^+ K^0$, $D^+_s \eta$. As can be seen in Refs.~\cite{Gamermann:2009uq,Song:2022kac}, $q_\text{max}$ provides the range of the interaction in momentum space. Consistently with the regularization of Eq.~\eqref{eq:Gcut}, the potential in Eq.~\eqref{eq:Vpot} implies a momentum dependence of the type $V(q, q') = V \theta(q_\text{max} - |{\bm q}|) \theta(q_\text{max} - |{\bm q'}|)$. The same model is used in Ref.~\cite{Liu:2023uly} to obtain the correlation functions. The interaction is the same except for the $V_{13}$, $V_{23}$ matrix elements. This is due to the fact that $\eta$ is implicitly assumed to be the SU(3) octet state in Ref.~\cite{Liu:2023uly} while we are using the $\eta$, $\eta'$ mixing of Ref.~\cite{Bramon:1994cb}.  This, however, does not have any effect in the $D^0 K^+$, $D^+ K^0$ correlation functions, since it is shown in Ref.~\cite{Liu:2023uly} that the $D^+_s \eta \to DK$ transition practically does not have any contribution to the $DK$ correlation functions. We also use another feature of the work of Ref.~\cite{Liu:2023uly} which is that due to the isospin violation, the $D_s^+ \pi^0$ channel also has a very small effect in the $DK$ correlation function, a fact that we use to neglect this channel in our approach, which simplifies the analysis at the bearable prize of neglecting the very small width of the $D^*_{s0}(2317)$ state.

\subsection{Correlation functions }
The correlation function for a given channel is given by
\begin{equation}
 C(p)=\int d^3 r S_{12}(r)|\psi(r, p)|^2 \,,
\end{equation}
where $S_{12}$ is a source function, accounting for the probability that a pair is formed at a certain relative distance $\bm{r}$. It is usually parametrized as,
\begin{equation}
S_{12}(r) = \frac{1}{(\sqrt{4\pi})^3 R^3} \exp\left(-\frac{r^2}{4R^2}\right) \, ,
\label{sec:source}
\end{equation}
and $R$ depends on the type of the reaction used, of the order of 1~fm in $p$-$p$ collision or 5~fm in heavy-ion collisions. $\psi(r,p)$ is the wave function of the pair of particles, where $p$ is the relative momentum, measuring the energy of the system as $\sqrt{s} =\omega_1 (p)+\omega_2(p)$

We follow the formalism of Koonin–Pratt \cite{Koonin:1977fh}, also followed in Ref.~\cite{Liu:2023uly}, a bit modified in Ref.~\cite{Vidana:2023olz} to account for the range of the interaction consistently with the use of the cut off regularization of the loop functions. The formulas of Ref.~\cite{Vidana:2023olz} are immediately translated to the present case and we can write the following correlation functions:

$D^0 K^+$ : 
\begin{align}
 C_{D^0K^+}(p_{K^+})=1+& 4\pi\,\int_0^{+\infty} drr^2 S_{12}(r)\ \theta(q_\text{max}-|p_{K^+}|) \nonumber \\
&\Big\{\big|j_0(p_{K^+} r)+T_{11}(\sqrt{s}) \, \widetilde{G}^{(1)}(s,r)\big|^2 \nonumber \\
&+ \omega_2 \big|T_{21}(\sqrt{s}) \, \widetilde{G}^{(2)}(s,r)\big|^2  \nonumber \\
&+ \omega_3 \big|T_{31}(\sqrt{s}) \, \widetilde{G}^{(3)}(s,r)\big|^2-j^2_0(p_{K^+}r)\Big\} \,,
\end{align} 
with
\begin{equation} 
p_{K^+} = \frac{\lambda^{1/2}(s,M_{D^0}^2,m^2_{K^+})}{2\sqrt{s}} \,.
\end{equation}

$D^+ K^0$ : 
\begin{align}
 C_{D^+K^0}(p_{K^0})=1+& 4\pi\,\int_0^{+\infty} drr^2 S_{12}(r)\ \theta(q_\text{max}-|p_{K^0}|) \nonumber \\
&\Big\{\big|j_0(p_{K^0} r)+T_{22}(\sqrt{s}) \, \widetilde{G}^{(2)}(s,r)\big|^2 \nonumber \\
&+ \omega_1 \big|T_{12}(\sqrt{s}) \, \widetilde{G}^{(1)}(s,r)\big|^2  \nonumber \\
&+ \omega_3 \big|T_{32}(\sqrt{s}) \, \widetilde{G}^{(3)}(s,r)\big|^2-j^2_0(p_{K^0}r)\Big\}\,,
\end{align} 
with
\begin{equation} 
p_{K^0} = \frac{\lambda^{1/2}(s,M_{D^+}^2,m^2_{K^0})}{2\sqrt{s}}\,.
\end{equation}

$D_s^+ \eta$ : 
\begin{align}
 C_{D_s \eta}(p_{\eta})=1+& 4\pi\,\int_0^{+\infty} drr^2 S_{12}(r)\ \theta(q_\text{max}-|p_{\eta}|) \nonumber \\
&\Big\{\big|j_0(p_{\eta} r)+T_{33}(\sqrt{s}) \, \widetilde{G}^{(3)}(s,r)\big|^2 \nonumber \\
&+ \omega_1 \big|T_{13}(\sqrt{s}) \, \widetilde{G}^{(1)}(s,r)\big|^2  \nonumber \\
&+ \omega_2 \big|T_{23}(\sqrt{s}) \, \widetilde{G}^{(2)}(s,r)\big|^2-j^2_0(p_{\eta}r)\Big\}\,,
\end{align} 
with
\begin{equation} 
p_{\eta} = \frac{\lambda^{1/2}(s,M_{D_s^+}^2,m^2_{\eta})}{2\sqrt{s}} \,.
\end{equation}

The quantities $\widetilde{G}^{(i)}$ are given by
\begin{align}
\widetilde{G}^{(i)}(s,r)&=\int \frac{d^3 q}{(2\pi)^3}\frac{\omega^{(i)}_1(q)+\omega^{(i)}_2(q)}{2\omega^{(i)}_1(q)\,\omega^{i}_2(q)}\nonumber \\
&\quad \cdot \frac{j_0(qr)}{s-\left[\omega_1^{(i)}(q)+\omega_2^{(i)}({q})\right]^2+i\epsilon}.
\label{eq:Gtilde}
\end{align} 

\subsection{Inverse problem}
Now we shall assume that we have the correlation functions provided by some experiment. In the absence of such data at present, we shall show how the method works by using the correlation functions generated using the model of section~\ref{sec:model}.

We shall make no assumption on the potential $V_{ij}$ except that for good reasons we take $V_{33} (3\equiv D_s \eta) = 0 $. This is the case for the local hidden gauge potential, but it is quite general since the vertex $\eta \eta V$ which will be induced in most models is zero. Thus, we have the potential (symmetric)
\begin{equation}
V=\left(\begin{array}{ccc}
V_{11} & V_{12} & V_{13} \\
& V_{22} & V_{23} \\
& & 0
\end{array}\right).
\end{equation}

Next, we will assume that the potential has isospin symmetry. Using the isospin wave function of Eq.~\eqref{eq:iso}, we impose that $\langle I=0 |V| I=1 \rangle = 0$ and conclude that
\begin{equation}
\begin{aligned}
& V_{11} = V_{22}, ~~~ V_{13} = V_{23},\\
& \langle DK, I=0 | V| DK, I=0 \rangle =V_{11}+V_{12} \, , \\
& \langle DK, I=1 | V| DK, I=1 \rangle =V_{11}-V_{12} \, .
\label{eq:Viso}
\end{aligned}
\end{equation}

Hence, our matrix $V$ gets a bit simplified to
\begin{equation}
V=\left(\begin{array}{ccc}
V_{11} & V_{12} & V_{13} \\
& V_{11} & V_{13} \\
& & 0
\end{array}\right) .
\end{equation} 

On the other hand, and in order to account for possible missing coupled channels, as well as for the possible contribution of a non-molecular state, we introduce energy dependent parts in the potential, as done in Ref.~\cite{Dai:2023cyo}, such that

\begin{equation}
\begin{aligned}
& V_{11}=V_{11}^{\prime}+\frac{\alpha}{M^2_V}(s-\bar{s}) \, , \\
& V_{12}=V_{12}^{\prime}+\frac{\beta}{M^2_V}(s-\bar{s})  \, , \\
& V_{13}=V_{13}^{\prime}+\frac{\gamma}{M^2_V}(s-\bar{s}) \, ,
\end{aligned}
\end{equation}
where $\bar s$ is the energy squared of the $D^0 K^+$ threshold. The $T$ matrix with the three coupled channels would be evaluated in Eq.~\eqref{eq:TBS} and we have now eight free parameters $V'_{11}$, $V'_{12}$, $V'_{13}$, $\alpha$, $\beta$, $\gamma$, $q_\text{max}$, and $R$. The question now is to see if from the knowledge of the three correlation functions we can determine these parameters and with which uncertainties. Once we have determined these parameters, we evaluate the magnitudes, $a_i$, $r_{0i}$ corresponding to the scattering length and the effective range
for each channel, determine if there is a bound state and, if so, we determine the couplings and the compositeness of the state, meaning the probabilities $P_i$ ($i=1,2,3$) for the three channels and their sum. 

The errors are evaluated directly on the magnitudes mentioned above, since the errors in the parameter are not too meaningful due to strong correlations between them. Indeed, from Eq.~\eqref{eq:Viso} we see that if we have a state with isospin $I=0$, what matters is $V_{11} + V_{12}$ and not the value of the individual parameters. The same can then be said about $\alpha$ and $\beta$ since what matters for $I=0$ is $\alpha+ \beta$. On the other hand, there is some correlation between $q_\text{max}$ and $V_{11}+ V_{12}$ since, taking a single channel, the $T$ matrix is
\begin{equation}
 T = \frac{1}{V^{-1} - G}.
\end{equation}
If we have a pole where $V^{-1} = G$, we can make a trade off between $q_\text{max}$ and $V$ such that $\delta V^{-1} = \delta G$ and then the pole appears at the same place [see discussion to this respect in Ref.~\cite{Dai:2023cyo}].

To determine the value of the observables and their uncertainty, we use the method of resampling~\cite{Press:1992zz,Efron:1986hys,Albaladejo:2016hae}. We take 31 points in each correlation function and an error of the value of the correlation function of 0.02, typical of present experimental correlation functions, then generate the centroid of the point with a Gaussian distribution probability and do so for all the 31 points of each correlation function. After that, we make a $\chi^2$ best fit to the data and determine the value of the parameters  $V'_{11}$, $V'_{12}$, $V'_{13}$, $\alpha$, $\beta$, $\gamma$, $q_\text{max}$, and $R$. Using these parameters, we calculate the different observables. We perform this procedure around 50 times and then determine the average value and dispersion of each one of the observables, scattering lengths, effective ranges, couplings and probabilities, and we also consider $R$ as an observable as well as the binding energy.

%%%%%%%%%%%%%%%%%%%%%%%%%%%%%%%%%%%%%%%%%%%%%%%%%%%%%%%%%%%%%%%%%%%%%%%%%%%%%%%%%%% 
\section{Results}
\begin{figure}[!t]
\begin{center}
\includegraphics[width=1.0\linewidth]{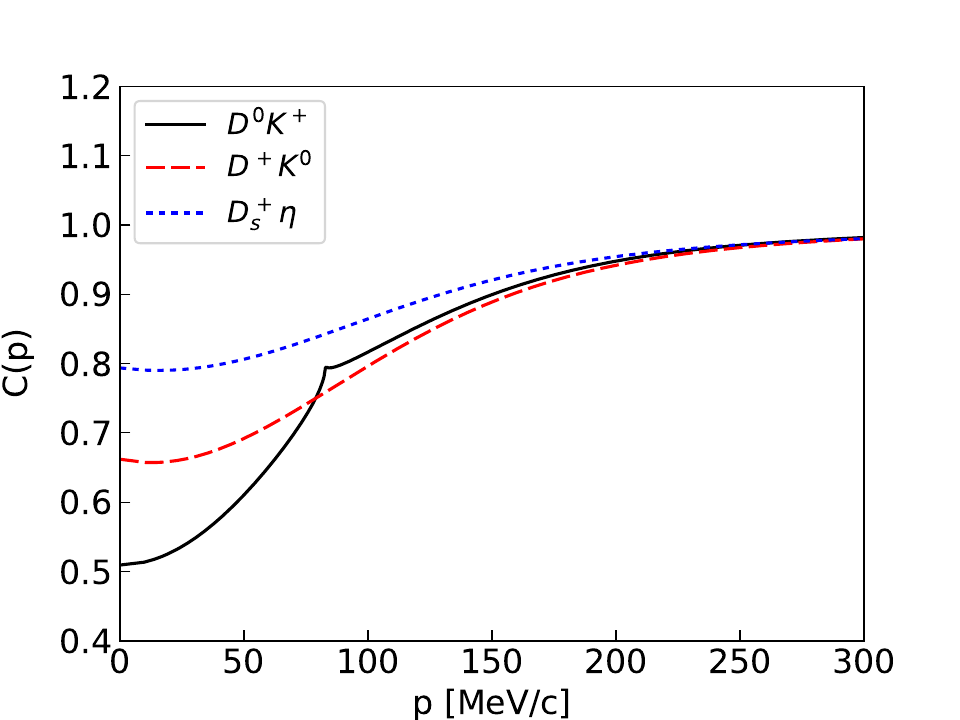}
\caption{ Correlation function of  $D^0 K^+$, $D^+ K^0$, and $D^+_s \eta$ calculated for $R=1$~fm.}
\label{fig:Cfunc}
\end{center}
\end{figure}

%%%%%%%%%%%%%%%%%%%%%%%%%%%%%%%%%%%%%%%%%%%%%%%%%%%%%%%%%%%%%%%%%%%%%%%%%%%%%%%%%%%
In Fig.~\ref{fig:Cfunc}, we show the result for the correlation function of  $D^0 K^+$, $D^+ K^0$, $D^+_s \eta$ calculated for $R=1$~fm. The correlation functions are very similar to those obtained in Refs.~\cite{Liu:2023uly} and \cite{Albaladejo:2023pzq} for the $D^0 K^+$, which are the only ones depicted in the papers. We have checked that the agreement also holds with the $D^+ K^0$ correlation function of Ref.~\cite{Liu:2023uly}\footnote{We thank Li-Sheng Geng for providing us with this information.}. In addition, we present the results for the $D_s^+ \eta$ correlation function. We find that the $D^0 K^+$ correlation function has bigger strength than the one for $D^+ K^0$ and the latter has bigger strength than that for the $D_s^+ \eta$ channel. We also corroborate the finding of 
Ref.~\cite{Liu:2023uly} that the effect of the $D_s^+ \eta$ channel in the other two correlation functions is negligible.

Although the precise values of the parameters are not very meaningful because of the strong correlations, we show the average value from all the fits for these parameters in order to get a feeling of their value. The $R$ parameter is uncorrelated and its uncertainty determines indeed the error of this parameter.
\begin{equation}
\begin{aligned}
 V_{11} &= -114.52 \pm 52.21 \,, \\
 V_{12} &= -117.31 \pm 54.53 \,, \\
 V_{13} &=  97.88  \pm 69.52 \,, \\
 \alpha &= -92.33  \pm 82.81 \,, \\
 \beta  &= -57.68  \pm 80.56 \,, \\
\gamma &= 35.06    \pm 81.40 \,, \\
 q_\text{max} &=  689.03 \pm 103.37 \ \text{MeV}\,, \\
 R &= 0.984 \pm 0.040 \ \text{fm} \,.
\end{aligned}
\end{equation}

From the values of the parameters of each fit of the resampling procedure, we can obtain $T_{ij}$ and from these different magnitudes: The binding energy of the resulting bound state, the scattering length $a_i$ and the effective range $r_{0,i}$ for the three channels. The coupling $g_i$ of the bound state obtained to each one of the channels and the probabilities $P_i$ for each channel in the wave function of the bound state obtained.

As  shown in Refs.~\cite{Gamermann:2009uq,Dai:2023cyo}, the $T$ matrix used here and the one in Quantum Mechanics have different normalization and we have for each channel
\begin{equation}
  T \equiv -8 \pi \sqrt{s} f^{Q M} \approx-8 \pi \sqrt{s} \frac{1}{-\frac{1}{a}+\frac{1}{2} r_0 k^2-i k}, 
\label{eq:T8pi}
\end{equation}
with
\begin{equation}
 k=\frac{\lambda^{1/2}\left(s, m_1^2, m_2^2\right)}{2 \sqrt{s}}.
\end{equation}
In one channel, we would have
\begin{equation}
 T = \frac{1}{V^{-1} - G}; \ \ T^{-1} = V^{-1} -G, 
\end{equation}
and 
\begin{equation}
 \text{Im} G = -\frac{1}{8 \pi \sqrt{s}} k.
\end{equation}
Then we can see that $ \text{Im} G $ cancels the $-i k$ term in Eq.~\eqref{eq:T8pi} and we can write
\begin{equation}
 -\frac{1}{a}+\frac{1}{2} r_0 k^2 \simeq-8 \pi \sqrt{s} T^{-1}+i k \,,
\label{eq:ar0}
\end{equation}
and $-8 \pi \sqrt{s}(-) \text{Im} G + i k=0 $. Eq.~\eqref{eq:ar0} holds for any of the three channels with $ T \equiv T_{ii}$, and from these we can determine $a_i$ and $r_{0,i}$ as
\begin{align}
 -\frac{1}{a} &=-8 \pi \sqrt{s} T^{-1} |_{s = s_\text{th}}, \\
 r_0 &=\frac{\partial}{\partial k^2}\ 2(-8 \pi \sqrt{s} \ T^{-1}+ik) \nonumber \\
& =\frac{\sqrt{s}}{\mu} \frac{\partial}{\partial s} \ 2(-8 \pi \sqrt{s} T^{-1} +i k) |_{s = s_\text{th}} \, ,
\end{align}
where $s_\text{th}$ is the squared of the energy of the system at threshold and $\mu$ is the the reduced mass of $m_1$ and $m_2$. For the $D^0 K^+$ channels, $a_1$ and $r_{0,1}$ will be real, but for the $D^+ K^0$ and $D^+_s \eta$ channels, these magnitudes are complex because the decays to the $D^0 K^+$ for $D^+ K^0$ or $D^0 K^+$, $D^+ K^0$ in the case of the $D_s^+ \eta$ channel are open. 

Taking into account that close to the peak of the bound state at $s_0$ we have
\begin{equation}
 T_{ij}  \sim \frac{g_i g_j}{s-s_0} \,,
\end{equation}
then we calculate the coupling as
\begin{equation}
 g_1^2=\lim _{s \rightarrow s_0}\left(s-s_0\right) T_{11} ; \quad g_j = g_1 \lim _{s\rightarrow s_0} \frac{T_{1 j}}{T_{11}},
\end{equation}
and the probabilities of each channel
\begin{equation}
P_i=- g_i^{2} \frac{\partial G_i}{\partial s} |_{s=s_0} .
\end{equation}

The results of the different fits described in the former section give us ($E$ stands for the energy of the bound state) 
\begin{equation}
\begin{aligned}
 E &= 2314.2 \pm 21.0 \ \text{MeV}, \\
 R &= 0.984 \pm 0.040 \ \text{fm} \,.
\label{eq:result.fit3}
\end{aligned}
\end{equation}
The values for the rest of the magnitudes are given in Table~\ref{tab:1}.

\begin{table*}
\caption{Values of the couplings, scattering lengths, effective ranges, and probabilities.} \label{tab:1}
\begin{tabular}{cccc}
\hline\noalign{\smallskip}
channel $i$ &~~~~~  $1: D^0 K^{+}$ & ~~~~~$2: D^{+} K^0$  &~~~~~ $ 3: D_s^{+} \eta$  \\
\noalign{\smallskip}\hline\noalign{\smallskip}
 $g_{i}$ [MeV] &~ ~~$8556.08 \pm 2707.16 $  & ~~$8571.21 \pm 2710.52$&~~~ $-6161.84 \pm 6307.93$   \\
 $P_{i}$  & ~~$0.357 \pm 0.133$  & ~~$0.306 \pm 0.119$&~~~ $0.083 \pm 0.070$   \\
 $a_{i}$ [fm]  & ~~ $0.720 \pm 0.131 $  & ~~$( 0.518 \pm 0.051) - i\,(0.120 \pm  0.030)$&~~~ $ ( 0.213 \pm 0.014) - i\,(0.054 \pm  0.025)  $   \\
 $r_{0,i}$ [fm] & ~~$-2.479 \pm 0.824$  & ~~$(- 0.162 \pm 0.778 ) -i\,(2.520 \pm  0.329)$&~~~ $( -0.165 \pm 1.677) -i \,( 0.171  \pm 0.663)$   \\
 \noalign{\smallskip}\hline
\end{tabular}
\end{table*}

\begin{table*}
\caption{Same as Table~\ref{tab:1} except with the two correlation functions of $D^0 K^+$ and $D^0 K^0$.} \label{tab:2}
\begin{tabular}{cccc}
\hline\noalign{\smallskip}
channel $i$ &~~~~~  $1: D^0 K^{+}$ & ~~~~~$2: D^{+} K^0$  &~~~~~ $ 3: D_s^{+} \eta$  \\
\noalign{\smallskip}\hline\noalign{\smallskip}
 $g_{i}$ [MeV] &~ ~~$7773.42 \pm 3462.55$  & ~~$7789.64 \pm 3483.53$&~~~ $-5716.45 \pm 5659.24$   \\
 $P_{i}$  & ~~$0.353 \pm 0.198$  & ~~$0.301 \pm 0.184$&~~~ $0.080 \pm 0.134$   \\
 $a_{i}$ [fm]  & ~~ $0.707 \pm 0.060 $  & ~~$( 0.504 \pm 0.034) - i\,(0.110 \pm  0.015)$&~~~ $ ( 0.259 \pm 0.067) - i\,(0.055 \pm  0.036)  $   \\
 $r_{0,i}$ [fm] & ~~$-3.139 \pm 1.299$  & ~~$(- 0.665 \pm 1.020 ) -i\,(2.386 \pm  0.341)$&~~~ $( 0.336 \pm 0.858) -i \,( 0.081  \pm 0.447)$   \\
 \noalign{\smallskip}\hline
\end{tabular}
\end{table*}

It is interesting to see that $P_1+P_2= 0.66 $ is not 1 indicating that there is a contribution from some other channels. In this case, we have also the $D_s^{+} \eta$  channel, and we see that $P_1+P_2+P_3$ is equal 0.75 within uncertainties, indicating that we have a largely molecular state. The result should not come as a surprise and is actually a tautology, since we started from a model to the correlation functions where the $D^*_{s0}(2317)$ is generated dynamically from the $D^0 K^{+}$, $D^{+} K^0$ and $D_s^{+} \eta$ channels with probabilities similar to these showing Eq.~\eqref{eq:result.fit3}. Note that $P_1 + P_2 \simeq 0.66 \pm 0.18$ also agrees with the results extracted from lattice QCD data in Ref.~\cite{MartinezTorres:2014kpc} of $0.72 \pm 0.13 \pm 0.05$.
What is a novelty of the present approach is that given some experimental correlation functions, the method exposed here provides all these magnitudes, and in particular the probabilities, with a fair accuracy. It is also interesting to see that the couplings $g_1$, $g_2$ are very similar, which indicates (see Ref.~\cite{Dai:2023cyo}) that according to Eq.~\eqref{eq:iso}, we have an $I=0$ state.

One good news from the present results is that from the correlations data one can obtain the value of $R$ with relative high precision. We started from the correlation functions using $R=1$~fm and the return of the fits is $ R=0.984 \pm 0.040$~fm, which means that we do not have to rely upon the experimental people to tell us the expected value of $R$ for the reaction that they use. The value of $R$ can be obtained from a fit to the data. 

It also should not be underestimated the fact that from the correlation functions which are above the threshold of the channels, one can induce the existence of a bound state related to the interaction of the particles. Once more we could obtain the binding energy with an uncertainty of about 20~MeV. It is worth noting here that in the study of lattice data in Ref.~\cite{MartinezTorres:2014kpc} the binding energy was determined with a similar uncertainty $B = 38 \pm 18 \pm 9$~MeV. 
We should also note that the value of the scattering lengths are obtained with relatively high accuracy but the effective ranges appear with large uncertainties.

We find it also interesting to see what happens if we perform a fit to the two correlation functions of $D^0 K^+$, $D^+ K^0$. Certainly we use less information and the uncertainties should be bigger. In order to see the value of using the three correlation functions, we compute only the binding and the probabilities, and find:

Fit to the $D^0 K^+$ and $D^+ K^0$ correlation functions,
\begin{equation}
\begin{aligned}
E &= 2322.2 \pm 19.5 \ \text{MeV}, \\
 % P_1 & = 0.353 \pm 0.198, \\
 % P_2 & = 0.301 \pm 0.184,  \\
 % P_3 & = 0.080 \pm 0.134, \\
 R &= 0.979 \pm 0.035 \ \text{fm} \,.
\label{eq:result.fit2}
\end{aligned}
\end{equation}

Comparison of the results of Eq.~\eqref{eq:result.fit2} and Table~\ref{tab:2} with those in Table~\ref{tab:1} obtained with three channels, tell us that we obtain a similar information with two channels as in three channels, only with a little bigger uncertainties. This information is useful from the experimental point of view since the $D_s^+ \eta$ channel might be more difficult to access. To complete the information provided in this case we show the values of the different magnitudes in Table~\ref{tab:2}.
We observe that the results for the magnitudes related to the $D^0 K^+$ and $D^+ K^0$ channels are similar to those in Table~\ref{tab:1}, done using three correlation functions, and the uncertainties are also similar. Only the magnitudes related to the $D_s \eta$ channel have now somewhat larger uncertainties. 

We should make a small remark concerning the fact that $P_1 + P_2 + P_3$ is not exactly one, although it can be close by counting uncertainties, in spite of using a model to generate the correlations that uses only the interaction between the molecular components. The reason has to be seen in the fact that the interaction of Eq.~\eqref{eq:p1p3p2p4}, stemming from the local hidden gauge approach, has some energy dependence and this has a consequence that $P_1 + P_2 + P_3 \neq 1$. For an energy independent potential the sum rule $P_1 + P_2 + P_3 = 1$ is exactly fulfilled~\cite{Gamermann:2006nm,Hyodo:2013nka}.
While this energy dependence has no practical effect in the determination of the probabilities for small binding energies, for large ones it produces effects attributing some probability to states outside those explicitly used in the approach.

\section{Summary}
We have investigated here the inverse problem of extracting information from the correlation functions in a model independent way. This approach is different to other methods used in the literature where a particular model is assumed and the parameters are tuned to get a good agreement with the data. We tested the approach with a problem related to the $D_{s0}^*(2317)$ resonance, which is tied to the interaction of the $D^0 K^+$, $D^+ K^0$, $D_s^+ \eta$ channels. In the absence of real data, we produced synthetic data from the correlation functions using a model based on the local hidden gauge approach for the interaction of mesons. The model, tuning a regulator parameter in the meson--meson loop functions tied to the range of the interaction, produces a bound state at 2317~MeV from the interaction of these three channels.

The inverse problem is based on one interaction, very flexible, which relies upon 8 parameters. 
We performed many fits to the data using the resampling method, and for each fit we evaluated the observables, obtaining from them the average value and the dispersion. The method is suited for a problem like the present one where we have a large correlation between the parameters. 
 This is one conclusion of our work, which is that the method used allows to show that different magnitudes tied to the interaction of the channels can be evaluated from the correlation functions with a relative accuracy.
The uncertainties obtained are similar to those obtained from present lattice QCD simulations~\cite{MartinezTorres:2014kpc}.
The precision obtained in scattering lengths and effective ranges is also similar to that obtained in the lattice calculation~\cite{MartinezTorres:2014kpc}\footnote{In Ref.~\cite{MartinezTorres:2014kpc}, the scattering lengths where extracted for the $I=0$ state, while here we evaluated them for each individual channel. Multiplying the result of Ref.~\cite{MartinezTorres:2014kpc} for $DK$ by $\frac{1}{\sqrt{2}}$, and changing the sign of $a$ ($\frac{1}{a}$ instead of $-\frac{1}{a}$ in Ref.~\cite{MartinezTorres:2014kpc}), the results are compatible within errors. The results of $r_0$ have large errors in both cases.}.

We also addressed the issue of the nature of the bound state obtained. Starting from a model that generates the bound state as a molecular state of the three considered channels, one expects to obtain the same results. This is the case here, but the interesting thing is that probabilities of the different channels can be obtained with some uncertainty that could be improved if the experimental errors are smaller than assumed here.
This is an important message that we can convey, since the analysis of correlation data along the lines discussed here can serve in the future to determine the nature of some bound states.

We also addressed the issue of how relevant is to use three correlation functions simultaneously, instead of two, and found that the use of the three correlation functions helped improve the accuracy in the determination of different magnitudes, but using just the $D^0 K^+$ and $D^+ K^0$ correlation functions already gave us enough information concerning these two channels. \\

\section*{Acknowledgments}
We acknowledge the hospitality of Universidad Nacional Autonoma de Mexico, where part of this work was carried out.
We thank Li-Sheng Geng for providing us with the result of the $D^+ K^0$ correlation function. 
N.~I. would like to thank Che Ming Ko for valuable discussions on the correlation function, and Kenta Itahashi, Biaogang Wu, and Pengsheng Wen for useful comments on the data fitting technique. 
The work of N. I. was partly supported by JSPS KAKENHI Grant Numbers JP19K14709 and JP21KK0244.
G.~T acknowledges the support from DGAPA-PAPIIT UNAM, grant no. IN110622.
This work is also partly supported by the Spanish Ministerio de Economia y Competitividad (MINECO) and European FEDER funds under Contracts No. FIS2017-84038-C2-1-P B, PID2020-112777GB-I00, and by Generalitat Valenciana under contract PROMETEO/2020/023. This project has received funding from the European Union Horizon 2020 research and innovation programme under the program H2020-INFRAIA-2018-1, grant agreement No. 824093 of the STRONG-2020 project.

%\clearpage

%\bibliographystyle{plain}
% \begin{thebibliography}{99}

% \end{thebibliography}

\bibliography{ref_Correlation}

\end{document}